\documentstyle[aps,multicol,psfig]{revtex}

\begin{document}
\title{POWER LAW STATISTICS OF AVALANCHES IN MARTENSITIC TRANSFORMATION} 
\author{Rajeev Ahluwalia$^{1,2}$ and G. Ananthakrishna$^{2,3}$}
\address{$^1$Jawaharlal Nehru Centre for Advanced Scientific Research,
Jakkur, Bangalore, India
$^2$ Materials Research Centre,
$^3$ Centre for Condensed Matter Theory, Indian Institute of Science,
Bangalore 560 012, India}
\maketitle
\begin{abstract}
We devise a two dimensional model that mimics the recently observed 
power law 
distributions for the amplitudes and durations of the acoustic emission 
signals observed during martensitic transformation 
[ Vives {\it et al}, Phys. Rev. Lett. {\bf 72}, 1694 (1994)]. 
We include a threshold mechanism arising from the athermal nature of 
transformation, long-range interaction between the transformed 
domains, inertial effects, and dissipation arising due to 
the motion of the interface. 
The model exhibits thermal hysteresis of the transformation, and
more importantly, it shows that the energy is released 
in the form of avalanches with power law distributions for their amplitudes 
and durations. Computer simulations also reveal 
morphological features similar to those observed in real systems.

\end{abstract}
\begin{multicols}{2}
Many spatially extended driven systems naturally evolve to 
a marginally stable state characterized by avalanches 
with power law distributions for their amplitudes and durations reflecting 
lack of intrinsic length scales and time scales in the system. 
 Such a state is termed 
as self-organized critical (SOC) state by Bak {\it et al }\cite{Bak}.
Several physical systems exhibit SOC features;
for example, earthquakes\cite{Rich}, 
acoustic emission from volcanic rocks\cite{Diod}, and stress drops
during  jerky flow \cite{Anan99}, to name a few.  
Recently, Vives {\it et al } \cite{Vives}, measured 
the acoustic  emission (AE) signals during martensitic transformation of 
Cu-Zn-Al single crystals under thermal cycling. 
They reported  power law statistics for the amplitudes and 
durations of the  AE signals both during cooling and heating runs. 
To the best of our knowledge, there is no strain (or 
displacement) based model of martensitic 
transformation which explains these results. More over,
 any prospective model has to 
take into account the non-equilibrium nature of the hysteresis.
Even though extensive theoretical studies exist on  
martensitic transformations\cite{Jdphys,Roit,Khacha}, the influence 
of dissipation and defects on hysteresis has received very little attention. 
Here, we propose a simple phenomenological model which captures 
the power law distribution of AE signals along with the thermal hysteresis 
of the transformation. Below, we will briefly collect 
SOC type dynamical features 
of the martensitic transformation relevant for modelling the system. 

Martensitic transformation is a
 first-order, solid-solid,  diffusionless, structural phase transition. 
On cooling, the unit cell gets distorted\cite{Jdphys,Roit,Khacha}
 leading to the nucleation of thin plate-like product domains 
with a twinned structure. This induces internal strains, which in turn 
induce long-range fields,
that block the transformation leaving the system in a two phase 
metastable state. ( Note that this implies the existence of 
built-in threshold mechanism.) Thus, the amount of 
the transformed phase is entirely determined by the excess free energy 
and, an additional undercooling 
is required for further growth. This implies that 
thermal fluctuations have little role in the transformation kinetics. 
Thus, the transformation is athermal and hence the
nucleation is athermal, usually occurring at defects like 
dislocations\cite{Jdphys,Khacha,Ferra,Cao}. 
Further, the emission of acoustic energy in the form of bursts implies 
that inertial effects are important. 
Indeed, Bales and Gooding, and later Reid and Gooding \cite{Bales}, 
have studied the importance of including the inertial effects. 
Vives {\it et al}  attribute the mechanism of 
irreversible release of the elastic energy in the form
 of avalanches, each of which correspond to the motion of one (or more) 
interface, to the  evolution of the system 
from one metastable state to another\cite{Vives}.  
Since the interface moves at near velocity of sound, 
there is dissipation associated with it. We include all these features 
along with the  long-range interaction between 
the transformed domains.

For the sake of simplicity, we consider, a {\it 2d} square to rectangle 
transition, for which the free energy is 
usually a function of all the three components of strain
defined by $e_{1}=(\eta_{xx}+\eta_{yy})/ {\sqrt{2}}, 
e_{2}=(\eta_{xx}-\eta_{yy})/ {\sqrt{2}}$,
and $ e_{3}=\eta_{xy}=\eta_{yx}$, where 
$\eta_{ij}={{1}\over{2}}({{\partial u_{i}}\over{\partial x_{j}}}
+{{\partial u_{j}}\over{\partial x_{i}}})$
refers to the components of the strain tensor and $u_i$   is the 
displacement field in the direction $i$ ( $i=x,y$). 
Here, $e_{1}$ is the bulk dilatational strain, $e_{2}$ is the 
 deviatoric strain and $e_{3}$ is the shear strain. 
In athermal martensites, 
deviatoric shear strains play a dominant 
role in the transformation kinetics\cite{Jdphys}. 
 In this paper, we consider the deviatoric strain $e_{2}=\epsilon(\vec{r})$ 
as the principal order parameter since 
volume changes are small \cite{Jdphys,Roit,Khacha,Jacobs}. The effect of 
other components of the strain, i.e., the bulk and
shear strain is accounted phenomenologically by considering a long-range 
interaction between the deviatoric strains. 

We write the rescaled free-energy  functional as 
$F\{\epsilon(\vec{r})\}=
{F_{L}}\{\epsilon(\vec{r})\} +  {F_{lr}}\{\epsilon(\vec{r})\},$
where $F_L$ is 
\begin{equation}
F_L  = \int d\vec{r}\bigg[f_l(\epsilon(\vec{r}))+
{D\over{2}}{{({\nabla}\epsilon(\vec{r}))^{2}}}
-\sigma(\vec{r})\epsilon(\vec{r})\bigg]. 
\end{equation}
where $D$ and $\sigma$ are in a scaled form. $F_{lr}$ is an effective long-range term that describes 
transformation induced strain-strain interactions. 
 In Eqn. (1), $f_l(\epsilon(\vec{r}))={{\delta T}\over{2}}{\epsilon(\vec{r})}^{2}
-{\epsilon(\vec{r})}^{4} + {{1}\over{2}}{\epsilon(\vec{r})}^{6}$ is the usual 
Landau polynomial for a first order transition, where $\delta T = (T - T_c)/(T_0 - T_c)$
 is the scaled temperature. $T_0$ is the first-order transition temperature 
at which the free energy for the product and parent 
phase are equal, and $T_c$ is the  temperature below which there are
only two degenerate global minima $\epsilon = \pm \epsilon_{M}$.
Following Cao {\it et al} \cite{Cao},
the effect of localized defects acting as 
nucleation sites is simulated by an inhomogeneous stress field, 
$\sigma(\vec{r})$. This term modifies the free-energy $f_l$ 
in a way that renders the austenitic phase locally
 unstable leading to the nucleation of the product phase.
As mentioned in the introduction, the physical cause of the
 long-range interaction is the coherency strain between the parent and 
the product phase. Such an interaction is also expected to arise due to 
the coupling of $\epsilon(\vec{r})$ with the other 
components of the strain order parameter.  Recently, an effective 
long-range interaction has been shown to result, both for the bulk \cite{Kartha,Shenoy}
 and interface \cite{Shenoy}, between the
deviatoric strains if $e_1$ and $e_3$, are eliminated by imposing
(St. Venant) elastic compatibility constraint. In ref \cite{Shenoy}, 
to describe the effect of the interface 
between the austenitic-martensite phases, the authors introduce 
an explicit interface term.
Instead, we follow Wang and Khachaturyan \cite{Wang} who have shown 
that the interface can be naturally described by 
accounting for coherency strains at the parent-product interface by 
a symmetry allowed fourth order anisotropic long-range interaction 
in the free energy. Following this, we write down the long-range  
term phenomenologically in the Fourier space as    
\begin{equation}
F_{lr}\{\epsilon\}=\int d\vec{k} B({\vec{k}}/k)
{\{ {\epsilon^{2}}(\vec{r})\}}_{k}
{\{ {\epsilon^{2}}(\vec{r})\}}_{k^{*}},
\end{equation}
where $\{ {\epsilon^{2}}(\vec{r})\}_{k}$ and ${\{ {\epsilon^{2}}(\vec{r})\}}_{k^{*}}$  
are the Fourier transform of $\epsilon^{2}(\vec{r})$ and its  complex 
conjugate respectively.  For the square to rectangle transformation, the 
favourable growth directions for the product phase are the habit plane 
directions $[11]$ and $[1\bar{1}]$. In addition, free-energy barriers should 
be large along the $[10]$ and $[01]$ directions.  
These features are well captured by the simple kernel 
$B({\vec{k}}/k)= -{{1}\over{2}}\beta \theta(k-\Lambda){{\hat{k}}^{2}}_{x}
{{\hat{k}}^{2}}_{y}$, where ${\hat{k}}_{x}$ and ${\hat{k}}_{y}$ 
are the unit vectors in the  $x$ and $y$ 
directions, and $\beta$ is the strength of interaction and
$\Lambda$ in the step function $\theta(k-\Lambda)$ is a cutoff 
on the interaction range. This kernel incorporates 
the effect of the interface in a natural way  since the cost of growth 
progressively increases with the transformation  in directions 
where the kernel is positive  which not only aids growth along the 
habit plane directions but also 
 limits the growth of domains transverse to it. 
We stress that this is only a simple choice and 
is not unique. Other kernels with similar 
orientation dependence will give similar results. 
 The real space picture of $B({\vec{k}}/k)$ is 
 similar to the long-range interaction of Kartha 
{\it et al} \cite{Kartha}. 

Even though deviatoric strain is the order parameter, basic variables 
are the displacement fields. Thus, we start with the 
Lagrangian $L=\mathcal{T}-F$,  where $F$ is the total free-energy 
with the kinetic energy $\mathcal{T}$ defined by 
\begin{equation}
{\mathcal{T}}=\int d\vec{r}\rho\bigg[\bigg({{\partial u_{x}(\vec{r},t)}\over{\partial t}}
\bigg)^{2}+
\bigg({{\partial u_{y}(\vec{r},t)}\over{\partial t}}\bigg)^{2}\bigg ].
\end{equation}
Here $\rho$ is the mass density. The dissipation associated with 
the movement of the parent-product 
interface is represented by 
the Rayleigh dissipative 
functional \cite{Landau}. 
Further, since deviatoric strains are the dominant ones, 
dissipative functional is written 
entirely in terms of $\epsilon(\vec{r})$,
 ie., $R={{1}\over{2}}\gamma\int d\vec{r}{\big({{\partial }\over{\partial t}}\epsilon
(\vec{r},t)\big)}^{2}$. (This is consistent with the fact that 
shear and bulk strains are known to equilibrate rapidly
and hence do not contribute.)
The equations of motion for the displacement fields are
\begin{equation}
{{d}\over{dt}}\bigg({{\delta L}\over{\delta {\dot{u}_i}}}\bigg)-
{{\delta L}\over{\delta u_{i}}}=-{ {\delta R}\over {\delta \dot{u}_i}}, \, \, i = x, y.
\end{equation}

\noindent
Using the above equations, we  obtain the equation  of motion for 
$\epsilon(\vec{r},t)$, which after scaling out $\rho$ and $D$ 
(in terms of rescaled space and time variables) gives
\begin{equation}
 \frac{\partial^2}{\partial t^2} \epsilon (\vec{r},t) =  \nabla^2 \bigg[ \frac{\delta F}{\delta \epsilon
(\vec{r},t)} + \gamma \frac{\partial}{\partial t} \epsilon (\vec{r},t)\bigg] \\
\end{equation}
\noindent
Thus, both $\beta$ and $\gamma$, here, are to be taken as rescaled parameters. 
The structure of Eqn. (5) is similar to that derived in \cite{Bales} for 1-d 
except for the  long-range term.

Simulations are carried out by discretizing Eq.(5) on a $N \times N$ grid
 using the Euler's scheme with periodic boundary conditions. 
The mesh size of the grid is $\Delta x=1$ and the 
time step $\Delta t=0.002$. 
The  long-range term is computed using psuedospectral method: this term is
calculated in the Fourier space and 
then inverse  fast Fourier transform 
is effected  to obtain the real space interaction. 
(We have used  the cutoff 
distance $\Lambda =0.2$.) 
 We  have studied both  the single defect 
and  many defect nucleation case. Here, we reports results on multi-site case 
and only mention results on single site case wherever necessary.

Consider the nucleation and growth 
for a single quench.
 We use a random distribution of defects and  represent their stress field 
by  $\sigma(\vec{r})={\sum_j^{j_{max}}}\sigma_0(\vec{r_j})
 exp [-|\vec{r}-\vec{r_j}|^{2}/\zeta^{2}]$,
where $\vec r_j$ refers to the scaled coordinates of the defect sites and 
$j_{max}$ is the total number of defect sites.
$ \sigma_0(\vec{r_j})$ is taken to be uniformly distributed in the 
interval $[-0.3,0.3]$. Initially, the system is taken to be in a homogeneous 
state with $\epsilon(\vec{r},0)$ distributed uniformly in the 
interval $[-0.005,0.005]$. At
$ t=0$, we  switch on the stress-field $\sigma(\vec{r}).$
Figure 1 shows snap shots of the transformation 
at $t= 10,\, 12,\, 15$ and $50$. (Grey regions represent the austenitic 
phase $\epsilon =0$, black and white regions  represent the two 
variants.)  Nucleation of the product phase occurs with a value 
$\epsilon \sim \pm 2$  at  several locations where the magnitude 
of the stress-field is sufficient to make the system locally 
unstable ( $t = 10$).  We note here that the  value of $\epsilon$ for 
the two variants is larger than that given by $f_l(\epsilon ({\vec r}))$ 
alone
( $ \epsilon_{M} \sim \pm 1.31$ for these parameters) due to the long-range interaction. 
In a short time, we see the other variant being created adjacent to these nuclei.
 By $t=12$ ( Fig. 1), the structure further 
develops into twinned arrays, propagating along $[11]$ and $[1\bar{1}]$ 
directions respectively. We also see that several new domains 
are nucleated {\it at a finite distance from the preexisting domains} 
which is consistent with the autocatalytic nucleation mechanism 
known to operate in martensitic transformation \cite{Khacha,Ferra}.
Though, the new nucleation sites,   most often coincide with the defect 
sites, occasionally it is seen in defect free regions.
( Similar observation was made in the single 
site case also.)    This can be attributed to the 
stress accumulation at these sites as a result of the long-range 
term due to the preexisting martensitic domains.  
Note that the twinning  is irregular which is again due to 
the mutual interaction between the various domains.
There is very little growth beyond  $t=30$. 
Thin needle-like structures can also be seen 
to emerge from larger domains ( $t = 50$) as reported by Vives et al.  
There is also a distribution of domains sizes.

We simulate thermal cycling of the transformation by continuously changing
 $\delta T$ from  + 40 to -80, and back in a duration of $t=1000$ units at a
constant rate, both for the heating and cooling runs.
For the reverse cycle, the final configuration of the cooling run 
is used as the initial configuration. 
Figure 2 shows the area fraction $\phi_A$  versus $\delta T$ 
for the cooling and heating runs ( $\circ$ ). 
In the cooling run, transformation starts around 
$\delta T_{ms} \sim - 2.0$ showing a rapid increase in $\phi_A$ up 
to  $\sim 30\%$, 
thereafter,  it increases linearly up to $90\%$.
 The transformation is complete  by $\delta T_{mf} \sim -59$. 
In the heating run, the parent phase appears only at 
$\delta T_{as} \sim -26.0$  and $\phi_A$ decreases almost linearly 
till the transformation is nearly complete around $\delta T_{af} \sim 18$. 
( For the sake of comparison, we have also shown the single-site hysteresis 
loop by $\bullet$. )

The {\it most important feature of the model is that 
the changes in} $\phi_A$ {\it are actually
jerky} which can only be seen on a finer scale. Since, the 
rate of energy dissipated  $ d E/d t = - 2 R(t)$, we have calculated $R(t)$ 
( or $R(\delta T)$). In Fig. 3, we have plotted $R(t)$ 
with the inset  showing the enlarged section of the peak which clearly 
shows  that the rate of energy release occurs in bursts 
consistent with the acoustic emission studies \cite{Vives}.
(Latent heat also shows a pattern similar to Fig. 3 \cite{Augu}.) 
Further, the distribution of the amplitudes of $R(t)$ 
denoted by  $R_A$ has a tendency to approach a power law 
$D_R(R_A)\sim R_A^{-\alpha_R}$ with $\alpha_R \sim 2.6$ ($\circ$ in Fig. 4. 
We have also shown the 
single-site results by $\bullet$.)  Similarly, 
the distribution $D_R(\Delta t)$ of the durations $\Delta t$ of 
the energy bursts scales with $\Delta t$ given by 
$D_R (\Delta t) \sim \Delta t^{ -\tau_R}$ 
with $\tau_R \sim 3.2$. 
Although, the scaling regime is almost identical to  $D_R(R_A)$, we find that 
the scatter is slightly more in this case. 
( This is true of experimental results as well\cite{Vives} and in SOC models\cite{Bak}).
The conditional average\cite{Rafols} of $R_A$, for a given $\Delta t$ denoted 
by  $< R_A>_c$ behaves as $ <R_A)>_c \sim \Delta t^{x_R}$. We get the 
exponent value $x_R \sim 1.36$. We also find that the scaling relation 
$ \tau_R = x_R(\alpha_R - 1) + 1 $ is satisfied quite well. 
In experiments, however, one measures the amplitude of the AE 
signals $A_{ae}$, ie., $R_A \sim A_{ae}^2$. Using the relationship between 
the two joint probability distributions 
$D(R_A, \Delta t) \propto D(A_{ae}, \Delta t)/A_{ae}$, it can be easily shown that  
$\alpha_R = (\alpha_{ae} + 1)/2$ where $\alpha_{ae}$ is the exponent 
corresponding to  $A_{ae}$. The other two exponents remain unchanged.
Using the experimental values\cite{Vives} ( $\alpha_{ae} \sim 3.8, \tau_{ae} 
\sim 3.6$ and $x_{ae} \sim 1$), we see that $ \alpha_R \sim 2.4$. Thus, 
we see that our values are in reasonable agreement with experiments 
considering the fact that real systems  are 3-d. 
It must be stated here that in 3-d even the number 
of variants are generally more and one also 
expects that the mechanisms operating in 3-d cannot 
be fully accounted for 
in 2-d\cite{Khacha,Wang}. We have also carried 
out a similar analysis on $R(t)$ for the  heating run. 
Even though, the changes in $R(t)$ occurs in bursts in
 the  heating run as well, we find that the scatter in the distributions is 
more than that for the cooling run.

In conclusion, the fact that the energy release occurs in the form of 
bursts with  a power law statistics  for the avalanches is well captured 
by the model. In addition, the model shows hysteresis  under thermal cycling. 
The power law statistics can be attributed to the fact that we have 
included important ingredients 
of SOC dynamics, namely, the threshold dynamics, 
dissipation, the generation of large number of metastable states 
during the transformation, and  a relaxation mechanism for the stored energy. 
This relaxation time scale corresponds to the inertial time scale as can 
be inferred from the fact that the basic  variables are the displacement fields 
which in turn set the limit on the fastest time scale. 
Indeed,  from our simulations  we find that the interface movement occurs at
time scales of a few units of (scaled) time.
Compared to this the driving force increases 
with temperature at a slow pace.  The  important feature of inducing large number 
metastable states during cooling or 
heating runs is due to the long-range interaction between the 
transformed domains as can be seen from the following reasoning. 
We note that the value of this term at any location is 
the result of the superposition of the 
contributions arising from the spatial distribution of the 
already transformed domains which in turn leads  
to a complex terrain of local  barriers ( metastable states). 
These {\it self generated} ( transformation induced) local thresholds,  
at a given time, must be overcome by 
the increase in the driving force arising from the 
slow cooling (or heating). We note that once a local barrier is 
overcome, part of the driving force goes in creating 
a new twin  and the rest is dissipated 
in the form of  burst of energy ( due to 
the advancing interface). The fact that long-range interaction 
is at the root of creating the local thresholds is {\it further supported by 
the fact that we find a power law distributions even in the single site 
nucleation case}. (See $\bullet$ in Fig. 4.) The presence of defect sites only 
triggers the initial 
nucleation process. This must be contrasted with disorder based Ising 
models\cite{Sethna93} which also produce power law statistics for avalaches
and field induced hysteresis. However, by subjecting the samples to repeated thermal cycling, 
Vives {\it et al} have verified that in martensite transformation, 
it is the dynamical ( transformation induced) disorder that is at the root 
of the avalanches. In this sense, {\it our model is the first to capture 
both thermal hysteresis and jerky  nature of the transformation based on 
dynamical disorder} and is independent of quenched disorder. 
Finally, the morphological features, including the needle shaped domains are 
very similar to those observed in experiments.

Acknowledgement: The authors thank Ms. M. S. Bharathi for 
the analysis of the data. Financial support of JNCASR 
for carrying out this project is gratefully acknowledged.

Figure captions:

Fig. 1: Snapshots of the morphology of the transformation at $t= 10,12,15$ 
and 50 time steps for multi-site nucleation. The parameters values are: 
$N = 128,\delta T = -2.0,\beta = 50, \gamma = 4, \sigma_0 = [-0.3, 0.3]$ and $j_{max} = 164$.

Fig. 2: Area fraction $\phi_A$ as a function $t$  for cooling 
and heating ($\circ$) runs for $N =256, 
\beta = 50, \gamma =4, \sigma_0 = [-0.3, 0.3] $ 
and $j_{max} \sim 1\%$ of $N^2 $ .
$\bullet$ correspond to single-site case.

Fig. 3: $R(t)$ as a function of $t$ in the cooling run 
for $N=256, \beta = 50, \gamma =4, \sigma_0 = [-0.3, 0.3]$ 
and $j_{max} \sim 1\%$ of $ N^2 $. The inset shows the fine structure of 
the peak.

Fig. 4: A plot of $D(R_A)$ versus $ R_A$ for the same parameter 
values as in Fig.3 for the multi-site case ($\circ$ connected by dashed line)
 with $\alpha = 2.6$. $\bullet$ with continuous line shows the single  
site case with $\alpha = 2.5$. 
\vspace{-0.3cm}
\begin{figure}
\centering
\mbox{
\psfig{figure=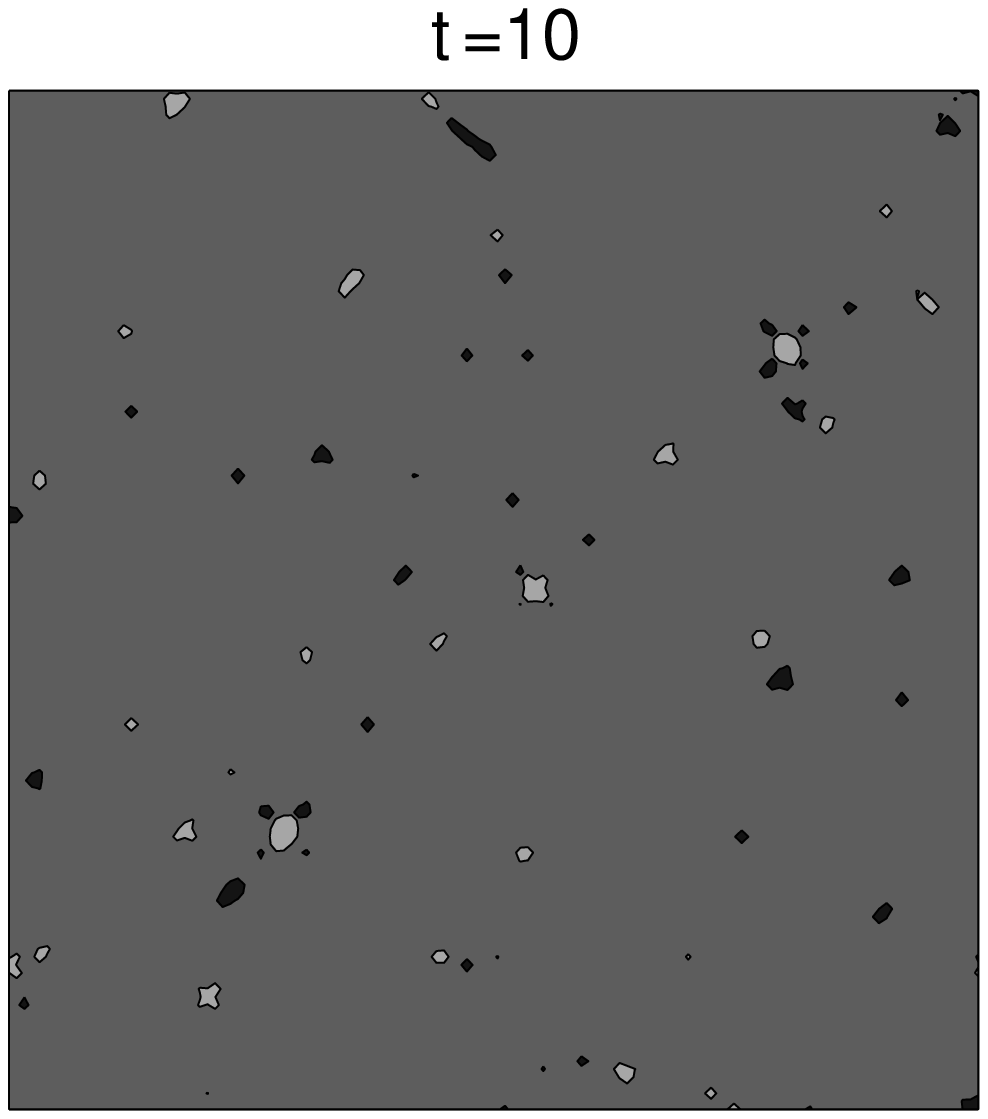,width=3cm}
\psfig{figure=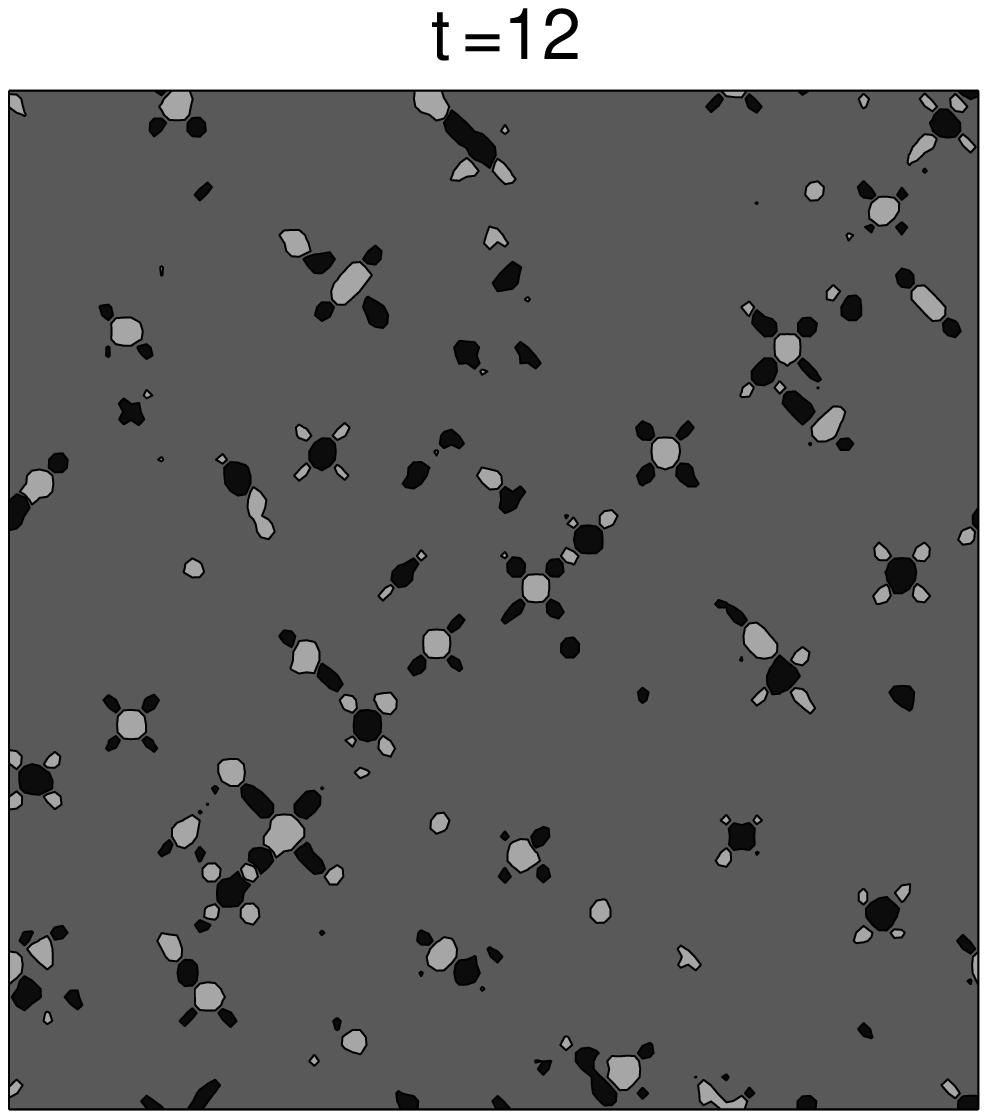,width=3cm}}
\end{figure}
\vspace*{-1cm}
\begin{figure}
\centering
\mbox{
\psfig{figure=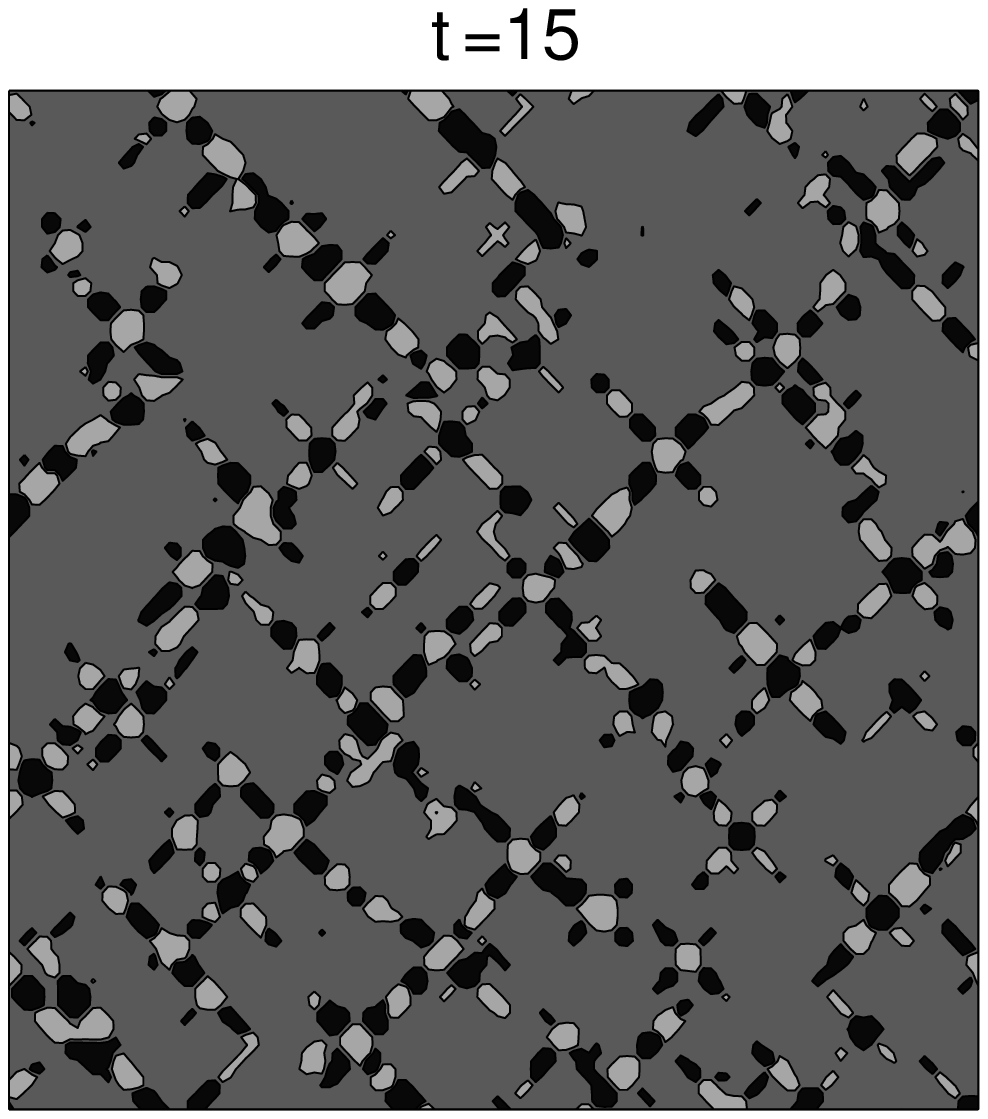,width=3cm}
\psfig{figure=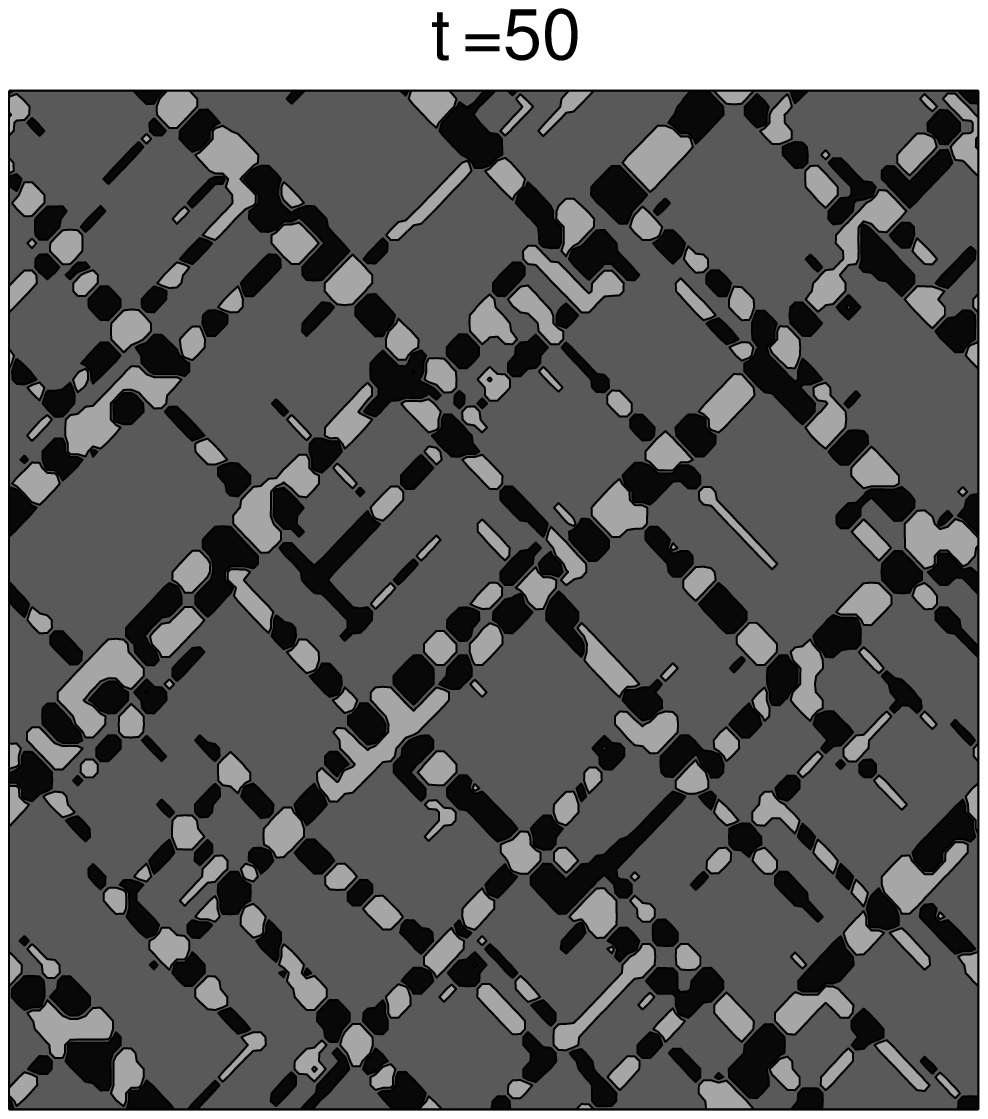,width=3cm}}
\end{figure}
\begin{figure}
\centerline{\psfig{file=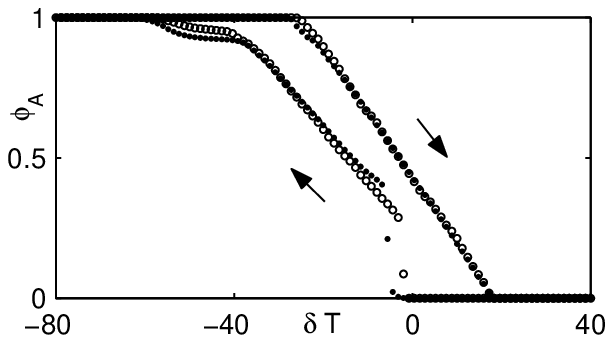,height=3.5cm,width=7cm}}
\end{figure}
\begin{figure}
\vspace{-0.5cm}
\centerline{\psfig{file=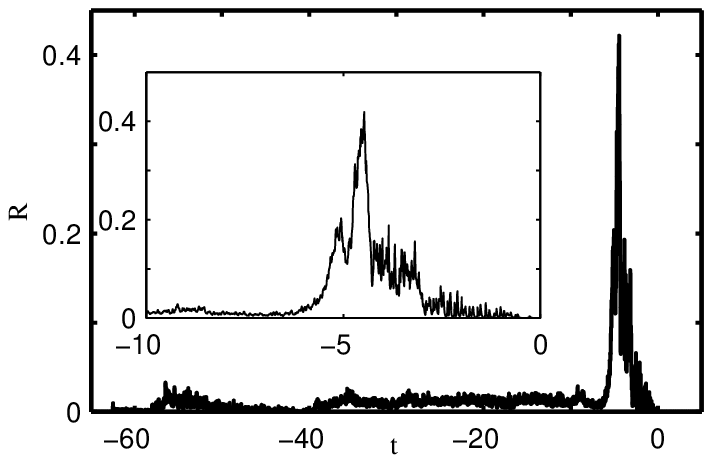,height=4cm,width=7cm}}
\end{figure}
\begin{figure}
\vspace{-0.5cm}
\centerline{\psfig{file=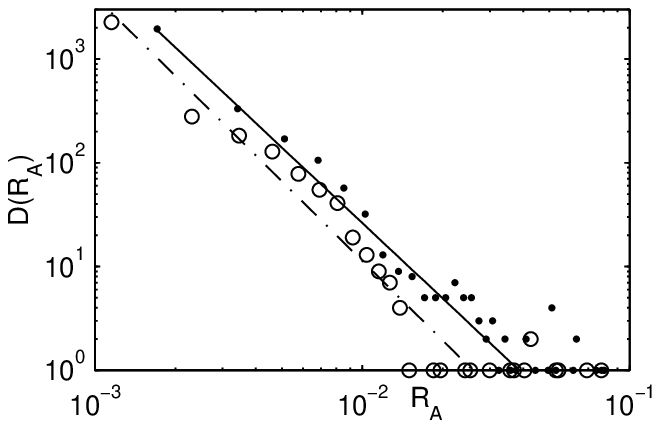,height=3.5cm,width=7cm}}
\end{figure}
\end{multicols}
\end{document}